\documentclass[prb,twocolumn,10pt,showpacs,nobibnotes,nofootinbib,floatfix,amsmath,amssymb]{revtex4}
\usepackage[T1]{fontenc}
\usepackage{mathptmx}
\usepackage{graphicx}
\usepackage{braket}
\usepackage{lettrine}

\newcommand{\total}{\operatorname{d}\!}
\newcommand{\varexp}{e}

\begin{document}
\title{Evaluation of High Order Terms for the Hubbard Model\\ in the Strong-coupling Limit}
\author{Eva Kalinowski}\email{kalinows@mathematik.uni-marburg.de}
\affiliation{Academy of Computer Science, 43-300 Bielsko-Bia\l a, Poland}
\author{W\l adys\l aw Gluza}
\affiliation{Silesian University of Technology, 44-100 Gliwice, Poland}
\date{24\textsuperscript{th} June 2011}

\begin{abstract}
\noindent The ground-state energy of the Hubbard model on a Bethe lattice with infinite connectivity at half filling is calculated for the insulating phase. Using Kohn's transformation to derive an effective Hamiltonian for the strong-coupling limit, the resulting class of diagrams is determined. We develop an algorithm for an algebraic evaluation of the contributions of high-order terms and check it by applying it to the Falicov--Kimball model that is exactly solvable. For the Hubbard model, the ground-state energy is exactly calculated up to order $t^{12}/U^{11}$. The results of the strong-coupling expansion deviate from numerical calculations as quantum Monte~Carlo (or density-matrix renormalization-group) by less than $0.13\,\text{\%}$ ($0.32\,\text{\%}$ respectively) for $U>4.76$.
\end{abstract}

\maketitle%

\section*{\MakeUppercase{Introduction}}

\lettrine{T}{he} Hubbard model\cite{Montorsi92} captures the essential elements of the complex behavior of strongly-correlated fermi\-onic systems with short-range repulsive interaction. Particularly interesting is the exploration of the transition from the paramagnetic metallic phase to a paramagnetic Mott--Hubbard insulator in the limit of infinite dimensions\cite{Gebhard97,Schlipf99,Rozen99,Krauth00,Kotliar00}; there, the interacting lattice problem can be mapped onto effective single-impurity models and solved within the framework of the dynamical mean-field theory (DMFT). Since this phenomenon was discussed controversly\cite{Eastwood03,Feldbacher04,Nishimoto04}, high accuracy in determining the ground-state energy and double occupancy per lattice site near the transition region is necessary for resolving doubts as to the nature of the transition, for minimizing quantitative uncertainties in the phase diagram and establishing an essential benchmark for other, in particular numerical methods. There were very many attempts to study the model in the strong-coupling limit (cf., e.\,g., refs.~\onlinecite{Chernyshev04,Delannoy05,Phillips04}). However, it appears to be rather difficult to go beyond the lowest orders. Therefore, we developed a computer-algebraic approach.

In this work we present a detailed description of a ``divide-and-conquer'' algorithm used for an exact calculation of all coefficients in the asymptotic expansion of the ground-state energy of the Mott insulator including order $t^{12}/U^{11}$. Results of such an algorithm up to $t^{10}/U^{9}$ were already quoted in ref.~\onlinecite{Bluemer04}, where an extrapolation scheme to infinite order (extrapolated Perturbation Theory, ``ePT'') was introduced. This showed excellent agreement with a quantum Monte~Carlo (QMC) technique, improved the state of the art by 1--2 orders of magnitude and lead to a well controlled evaluation of the critical exponent. Quite recently, our method was applied to the Bose--Hubbard model\cite{Teichmann09,Eckardt09}.

The outline of this paper is as follows: In sec.~I, we show how the effective Hamiltonian is derived following Kohn's\cite{Kohn64} and Kato's\cite{Kohn64} and Takahashi's\cite{Takahashi77} treatment of the strong-coupling limit. Then, in sec.~II, the class of diagrams defined by the resulting effective Hamiltonian is discussed for the Bethe lattice with infinite connectivity, and the algorithm for the evaluation of electronic transfer processes on it is described. The concept of this algorithm is ideally suited for parallelization that will be done in further work. The results are first given for the Falicov--Kimball model that is exactly solvable and serves as a test of our treatment (sec.~III.A). Our main result is given in eqs.~(\ref{eqn:HubbardEnergy}) and~(\ref{eqn:HubbardDouble}) in sec.~III.B. We apply our method ``ePT'' (see ref.~\onlinecite{Bluemer04}) and compare our results with results from DMFT-QMC and DMFT-DDMRG (Dynamical Density-Matrix Renormalization Group)\cite{Nishimoto04} for the ground-state insulating phase of the Hubbard model. Finally, flowcharts that present the essential parts of the algorithm are given in the Appendix.

\section{Perturbation Expansion for the Strong-coupling Limit}

\noindent We investigate spin-$1/2$ electrons on a lattice represented by the Hubbard model
\begin{equation}\label{eqn:Hubby}
  H=T+UD\,\text{,}
\end{equation}
where $T=-t\sum_{(\mathbf{i},\mathbf{j}),\sigma}c^{\dagger}_{\mathbf{i}\sigma}c^{\text{}}_{\mathbf{j}\sigma}$ is the kinetic energy operator describing electron hops between nearest neighbour sites~$\mathbf{i}$ and~$\mathbf{j}$ with the transfer amplitude $t$, $UD=U\sum_{\mathbf{i}}n_{\mathbf{i}\uparrow}n_{\mathbf{i}\downarrow}$ is the interaction part including only local contributions $n_{\mathbf{i}\sigma} = c_{\mathbf{i}\sigma}^{\dagger}c_{\mathbf{i}\sigma}^{}$. $c_{\mathbf{i}\sigma}^{\dagger}$ and $c_{\mathbf{i}\sigma}^{}$ are the creation and annihilation operators for electrons with spin $\sigma=\downarrow,\uparrow$ on site $\mathbf{i}$. 

In the following, we sketch the calculation of an effective Hamiltonian in a strong coupling expansion, as was developed in ref.~\onlinecite{Kato49} and ref.~\onlinecite{Takahashi77}. There it is shown how this expansion in $1/U$ is done systematically. The aim is the transformation to new particles with an effective Hamiltonian that does not change the number of doubly occupied sites. The considerations are valid for any lattice in any dimension. The operator for the kinetic energy $T$ in eq.~(\ref{eqn:Hubby}) couples states with a different number of doubly occupied sites. In deriving this effective Hamiltonian, a decoupling can be achieved by introducing suitable linear combinations. Rotating to such a new basis is performed by a unitary transformation $\mathcal{U}\equiv\varexp^{S}$ developed by Kohn\cite{Kohn64} for the strong-coupling limit. This transformation introduces new particles created by $\tilde c^{\dagger}$ so that
\begin{equation}\label{eqn:Trafo}
c^{\dagger}_{\mathbf{i}\sigma}=\varexp^{S(\tilde c)}\tilde c^{\dagger}_{\mathbf{i}\sigma}\varexp^{-S(\tilde c^{})}\,\text{,}
\end{equation}
and therewith
\begin{equation}\label{eqn:Transformation}
H(c)=\varexp^{S(\tilde c)}H(\tilde c)\varexp^{-S(\tilde c)}\equiv\tilde H(\tilde c)\text{.}
\end{equation}
The generator $S$ is constructed in such a way that the hopping of the new particles does not change the effective number of doubly occupied sites for the new particles ($\tilde{c}$), 
\begin{equation}\label{eqn:Kommutator}
[\tilde H(\tilde c),D(\tilde c)]=0\,\text{.}
\end{equation}
This requires $S$ (and therefore $\tilde H(\tilde c)$) to be an operator series in $1/U$
\begin{equation}\label{eqn:SReihe}
S(\tilde c)=\sum_{i=1}^{\infty}\frac{S_i(\tilde c)}{U^i}\,\text{,}
\end{equation}
and the unitarity of the transformation implies $S^{\dagger}=-S$. Obviously, the wavefunctions can be expressed in terms of the new particles, and it follows for the eigenenergies
\begin{equation}
\langle\psi_m(c)|H(c)|\psi_m(c)\rangle=\langle\tilde\psi_m(\tilde c)|\tilde H(\tilde c)|\tilde\psi_m(\tilde c)\rangle=E_m\,\text{.}
\end{equation}
The ground state $\tilde\psi_0(\tilde c)$ of $\tilde H(\tilde c)$ at half band filling will be determined at the end of this section.

The low orders in $1/U$ are conveniantly found by substituting the expansion (\ref{eqn:SReihe}) in (\ref{eqn:Transformation}); to second order in $1/U$:
\begin{multline}\label{eqn:EffOperator}
\tilde H(\tilde c)=T(\tilde c)+UD(\tilde c)+\frac{1}{U}[S_1(\tilde c),H(\tilde c)]+\\+\frac{1}{U^2}[S_2(\tilde c),H(\tilde c)]+\frac{1}{2U^2}[S_1(\tilde c),[S_1(\tilde c),H(\tilde c)]]+\cdots\,\text{.}
\end{multline}
From the condition (\ref{eqn:Kommutator}), the coefficients~$S_n(\tilde c)$ are determined order by order as shown now. The kinetic energy operator can be separated in three parts, each of which increases or decreases the number of double occupancies by one, or leaves it unchanged,
\begin{equation}
T(\tilde c)=T_{U}+T_{-U}+T_0\,\text{,}
\end{equation}
where
\begin{align*}
T_{U}&=-t\sum_{(\mathbf{i},\mathbf{j}),\sigma}\tilde n_{\mathbf{i},-\sigma}(1-\tilde n_{\mathbf{j},-\sigma})\tilde c^{\dagger}_{\mathbf{i}\sigma}\tilde c^{}_{\mathbf{j}\sigma}\,\text{,}\\
T_{-U}&=-t\sum_{(\mathbf{i},\mathbf{j}),\sigma}\tilde n_{\mathbf{j},-\sigma}(1-\tilde n_{\mathbf{i},-\sigma})\tilde c^{\dagger}_{\mathbf{i}\sigma}\tilde c^{}_{\mathbf{j}\sigma}\,\text{,}\\
T_{0}&=-t\sum_{(\mathbf{i},\mathbf{j}),\sigma}(1-\tilde n_{\mathbf{i},-\sigma}-\tilde n_{\mathbf{j},-\sigma}+2\tilde n_{\mathbf{i},-\sigma}\tilde n_{\mathbf{j},-\sigma})\tilde c^{\dagger}_{\mathbf{i}\sigma}\tilde c^{}_{\mathbf{j}\sigma}\,\text{.}
\end{align*}
Because
\begin{equation}
[T_{U},D(\tilde c)]=-T_{U}\quad \text{and}\quad [T_{-U},D(\tilde c)]=T_{-U}\,\text{,}
\end{equation}
(\ref{eqn:Kommutator}) is fulfilled when the cross terms $T_{U}$ and $T_{-U}$ are cancelled by $[S(\tilde c),UD(\tilde c)]$ in the lowest order in $U^{-1}$. This is achieved by choosing 
\begin{equation}
S_1(\tilde c)=T_U-T_{-U}\,\text{.}
\end{equation}
Inserting in (\ref{eqn:EffOperator}) and demanding (\ref{eqn:Kommutator}), one obtains the condition for the next order, i.\,e., $S_2$ that leads to
\begin{equation}
S_2(\tilde c)=[T_U+T_{-U},T_0]\,\text{.}
\end{equation} 
Following this procedure, one determines $S(\tilde c)$ order by order. Since $S$ does not create or annihilate bare particles, the vacuum state is equal for both old and new particles. 

The lowest order terms of the resulting $1/U$-expansion of the Hamiltonian $\tilde H(\tilde c)=UD(\tilde c)+\sum_{i=0}^{\infty}U^{-i}\tilde h_i$ are (here and in the following, $T=T(\tilde c)$)
\begin{subequations}\label{eqn:ExpandedOp}
\begin{align}
\tilde h_0&=\sum_{j}P_jTP_j\,\text{,}\\
\tilde h_1&=\sum_{j}P_jTS^1_jTP_j\,\text{,}\\
\begin{split}
\tilde h_2&=\sum_{j}P_jTS^1_jTS^1_jTP_j-\\&\quad-\frac{1}{2}\sum_{j}\left(P_jTS^2_jTP_jTP_j+P_jTP_jTS^2_jTP_j\right)\,\text{,}
\end{split}\\
&\vdots\notag
\end{align}
\end{subequations}
where $P_j$ projects onto the subspace with $j\geq 0$ double occupancies and $S_j^k$ is defined as ($D_j=j$)
\begin{equation}\label{eqn:esika}
S^k_j =\sum\limits_{l\neq j}^{}\frac{P_l}{(D_j-D_l)^k}\,\text{,}\quad k>0\,\text{.}
\end{equation}
Calculation of the next orders in this way needs an increasing computational effort. Therefore, a computer program has been developed that evaluates the general formula of Kato, ref.~\onlinecite{Kato49} (cf.~eq.~22), up to a given order.

$\tilde H(\tilde c)$, the first terms of which are given by \eqref{eqn:ExpandedOp} is the desired effective Hamiltonian. It is valid in any subspace with fixed number of $j$ double occupancies (of particles corresponding to $\tilde c^{\dagger}$).
For large $U$, the ground state $\tilde\psi_0(\tilde c)$ of $\tilde H(\tilde c)$ must have the lowest value of $UD(\tilde c)$, and because $\tilde H(\tilde c)$ leaves the number of doubly occupied sites unchanged, see (\ref{eqn:Kommutator}), this state does not contain any double occupancies at half band filling, so we put $j=0$ in the following. So far, the considerations are valid for arbitrary dimension. Now, we focus on the case of half-filling in infinite dimensions. Then, all global singlets are ground states of $\tilde H(\tilde c)$, cf.~ref.~\onlinecite{KalinowskiThesis}. Therefore, each lattice site is equally likely to be occupied by an electron with spin $\uparrow$ or $\downarrow$, irrespective of the spin on any other lattice site. This enables us to perform the ground-state expectation values in $\braket{\tilde{\psi}_{0}(\tilde{c})|\tilde{h}_{n}|\tilde{\psi}_{0}(\tilde{c})}$ in the case of a half-filled band in a second computer program.

Both these computer programs are of algebraic nature (work with integers only) and thus give exact results for any given order in $1/U$.

\section{Graphs and Algorithm}

\begin{figure}[t]
  \includegraphics{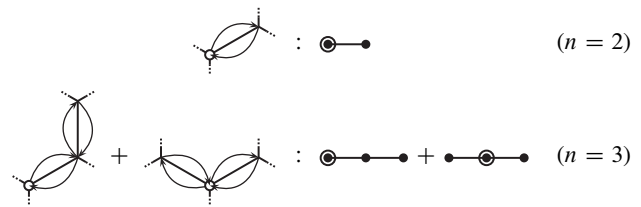}
  \caption{\label{fig:TheGraphs}Correspondence between sequences of hops on the lattice (left) and Butcher trees (right): Shown are the sequences in first ($n=2$) and third order ($n=3$) and the related Butcher trees. The first hop defines the root of the tree (encirceled).}
\end{figure}

\noindent In this section, we give more details regarding the evaluation of $\tilde H(\tilde c)$. In fixed order~$i$, the operator $\tilde h_i$ containes all possible electron hops resulting from $i+1$ applications of $T$; it contributes to the energy in the order $1/U^i$. In the following, we will deal only with states with zero doubly occupied sites (in the new particles), so we drop the index~$0$ and denote $S_0^k\equiv S^k$ and $P_{0}\equiv P$. A sequence of electron hops described through an operator chain $PTS^{k_1}\cdots S^{k_i}TP$ is called \emph{process}\cite{KalinowskiThesis}. Only processes that restore the initial state contribute to the ground-state energy. Consider now a given process and perform the sum on lattice sites in the operators~$T$. The individual terms in this sum are called ``sequences'' or ``diagrams''. Because of the linked-cluster theorem, we need to keep only connected diagrams. Each of these contains $n=2+\frac{i-1}{2}$ sites, connected through $i+1$ jumps. Diagrams of odd order in $t$ do not contribute at half filling for any lattice type. Now, we specialize our considerations to the Bethe lattice. There, all closed paths are self-retracing. This can be seen in fig.~\ref{fig:TheGraphs} where the possible sequences (of hops) for the low orders $i=1$ ($n=2$) and $i=3$ ($n=3$) are displayed on a Bethe lattice of connectivity~$3$. In the following, we put $t=t^{\ast}/\sqrt{Z}$ ($Z$ is the number of nearest neighbours) and consider the limit~$Z\to\infty$ for fixed $t^{\ast}\equiv 1$. This limit implies that in the energy, in each order in $U^{-i}$, the leading order $t^{i+1}$ is taken into account. Thus, diagrams with more than two transitions between any two given sites are suppressed at least as $1/Z$: every additional connection of already doubly connected sites is smaller by $1/Z$ compared to those with only two jumps between any two sites. Since the paths are self-retracing, they can be collapsed into `Butcher trees'\cite{Butcher76} as also indicated in the right of fig.~\ref{fig:TheGraphs}. The position of the first hop in the sequence (the index $\mathbf{j}$ of $c_{\mathbf{j},\sigma}$ at the rightmost $T$) defines the ``root'' of the tree, cf.~fig.~\ref{fig:TheGraphs}. Because there are exactly two hops between neighboring sites ($Z\to\infty$), there is a one-to-one correspondence between sequences and Butcher trees. The number of Butcher trees built with $n$ vertices, $A[n]$, is still moderate for moderate $n$; it is given, see ref.~\onlinecite{A000081}, by the following recursive definition
\begin{equation}
A[n]:=\frac{1}{n-1}\sum\limits_{j=1}^{n-1}A[n-j]\sum\limits_{k=1}^{k_{\text{max}}}\big(d_k(j)A[d_k(j)]\big)\,\text{,}
\end{equation}
with $A[1]:=1$ and $A[2]:=1$, and $d_k(j)$ is the $k$\textsuperscript{th} element of the set of divisors of $j$. In table \ref{tab:ButcherCount}, we give the number of trees, the number of initial spin configurations, and the number of sequences for given order of perturbation theory.

\begin{table}[t]
\begin{tabular}{c|c|c|c|c}
$i$ & $n$ & $A[n]$ & $B[n]$ & $C[n]$ \\
\hline%
$1$ & $2$ & $1$ & $2$ & $2$ \\
$3$ & $3$ & $2$ & $8$ & $20$ \\
$5$ & $4$ & $4$ & $32$ & $648$ \\
$7$ & $5$ & $9$ & $136$ & $45472$ \\
$9$ & $6$ & $20$ & $596$ & $5644880$ \\
$11$ & $7$ & $48$ & $2712$ & $1099056000$ \\
$\vdots$ & $\vdots$ & $\vdots$ & $\vdots$ & $\vdots$
\end{tabular}
\caption{\label{tab:ButcherCount}Number of trees with $n=(i+3)/2$ sites or order $i$ in perturbation theory, $A[n]$, see text. $B[n]$ is the number of the initial spin configurations; $C[n]$ is the number of all hopping sequences, i.\,e.\ the number of different ways to realize all processes of the given order with initial spin configurations.}
\end{table}

\begin{figure*}
  \includegraphics{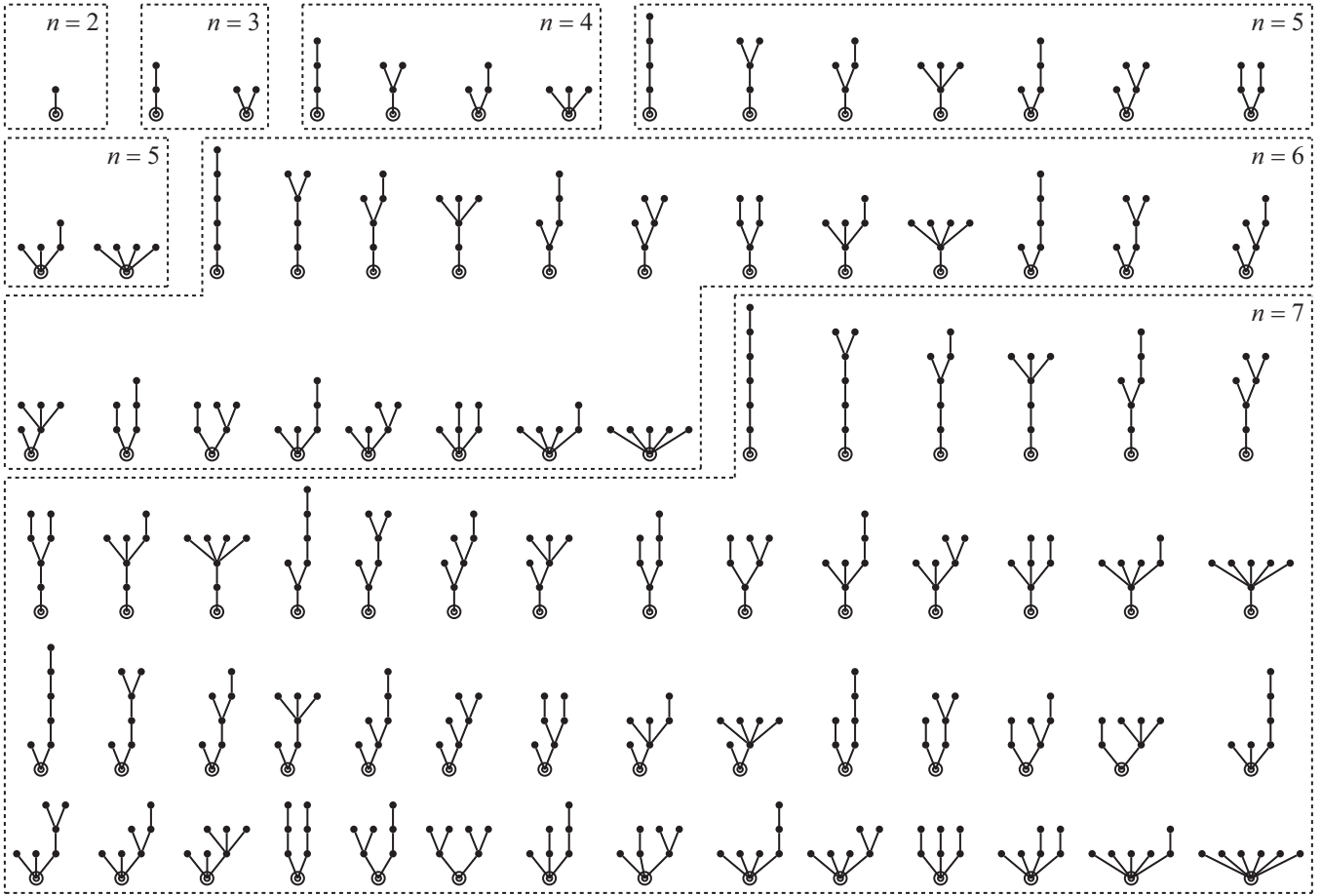}\medskip\par%
  \caption{\label{fig:ButcherTrees}Butcher trees up to the order $i=11$. The order expressed by the number of sites is $i=2n-3$. The encirceled sites denote the root of the trees.}
\end{figure*}

Illustrating the complexity of the problem, figure \ref{fig:ButcherTrees} shows the Butcher trees up to seven vertices (representing the connected sites) and thus all graphs contributing up to eleventh order in the perturbation series.

\begin{figure*}[htb]
  \includegraphics{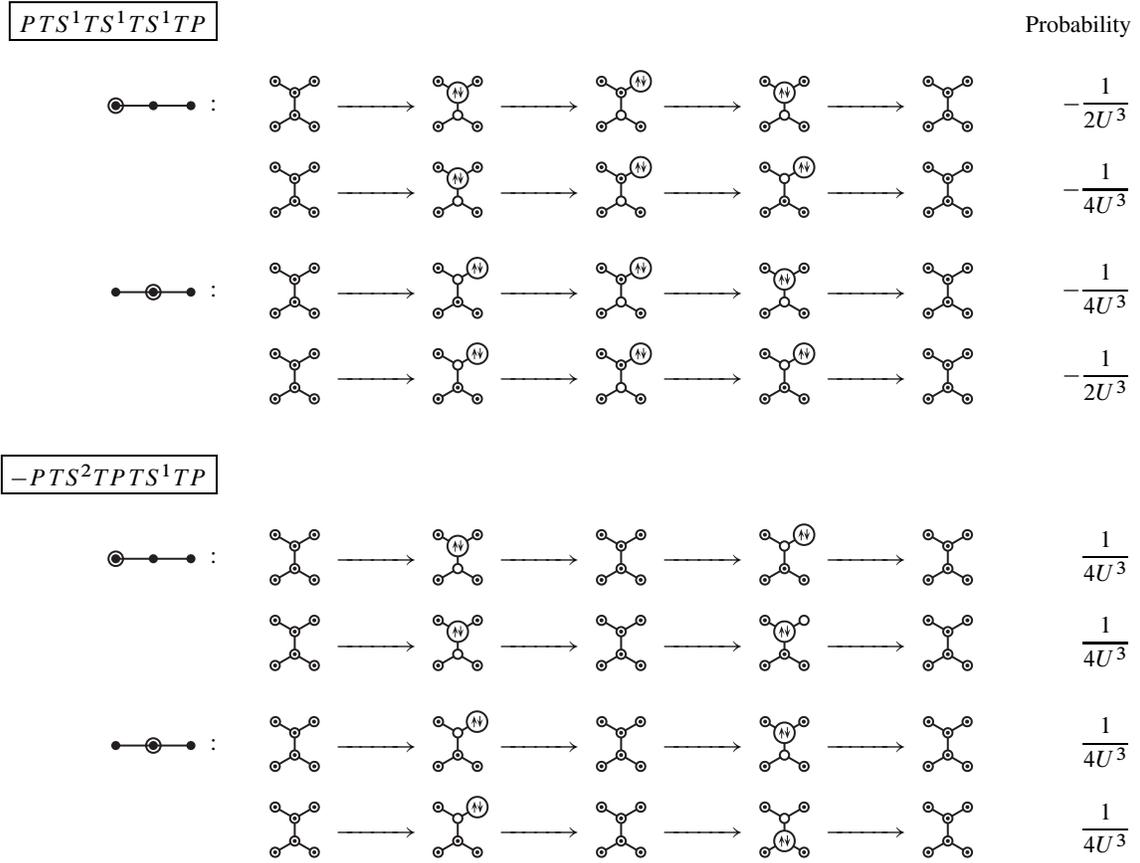}\smallskip\par%
  \caption{\label{fig:ProcessThirdOrder}The intermediate states for the two processes and all sequences contributing to the ground-state energy in third order in $\frac{1}{U}$. The arrows correspond to the application of $T$. The symbol \textcircled{$\scriptstyle\uparrow\!\downarrow$} denotes a doubly occupied lattice site; the symbols \textcircled{} and \textcircled{$\cdot$} denote a hole and a singly occupied lattice site, respectively. In the right column, the contribution of each sequence to the ground-state energy is indicated. They sum up to $-\frac{1}{2U^{3}}$.}
\end{figure*}

In order to illustrate our procedure, we show all intermediate states for all sequences for all processes in third order in fig.~\ref{fig:ProcessThirdOrder}. In the main part of the figure, these states are displayed on a Bethe lattice with connectivity~$3$. Note that only three sites are affected by the hops, the two in the center and the upper right one on this segment of the lattice. All other sites of the lattice are unaffected by the hops and we can restrict our considerations to the ``Butcher trees'' shown in the left part of the figure. There, the sites of the first hop, the roots of the trees, are encircled.
The processes are generated by following parts of the hamiltonian (terms containing the sequence $\dots PTP\dots$ vanish at  half band filling (hf)):
\begin{equation}
  \tilde h_{3}^{\text{hf}}=PTS^1TS^1TS^1TP-PTS^2TPTS^1TP\,\text{.}
\end{equation}
This expression coincides with the fourth order (highest available) in ref.~\onlinecite{Chernyshev04}, eq.~(6). The probabilities for the occurrence of the processes in the paramagnetic phase, multiplied by the prefactors of the processes in $\tilde h_{3}^{\text{hf}}$ ($1$ and $-1$ here), are given in the right column of figure \ref{fig:ProcessThirdOrder}. Their sum yields the contribution to the ground-state energy. 

The numerical algorithm to calculate the expectation value of the operators $\tilde h_i$ is based on this diagrammatic approach: After constructing all $i$\textsuperscript{th} order Butcher trees for the lattice, all possible sequences on them resulting from different terms of the $\tilde h_{i}$ are generated through a recursive procedure, within which the conditions for the realisation of the electron transfers, as the fulfillment of the Pauli principle and the consideration of preceding hopping steps on the branches, are defined. The first electron hop starts from the root of the graph (encircled sites in figures \ref{fig:ButcherTrees} and \ref{fig:TheGraphs}) to a neighbour site. In a single loop of the program, possible following jumps are tested by a subroutine and executed where applicable. Thereby, the second and last electron hop on a branch has to be performed by the same spin species as the first one. This requirement guarantees the restoration of the initial spin configuration. The actual number of double occupancies that enters the operators $S^{k}$, (\ref{eqn:esika}), is stored and used for the computation of the factor for the given process. The final spin configuration determines the factor's sign, $(-1)^P$, where $P$ is the number of permuted spin pairs. Summation yields the contribution of given order to the ground-state energy of the Mott insulator.

As shown now, this algorithm has been successfully tested by computing the ground-state energy of the exactly solvable Falicov--Kimball model, a simplified Hubbard model with one immobile spin species.

\section{Results}

\subsection{Falicov--Kimball Model}

\noindent For the Hamiltonian of the half-filled Falicov--Kimball model we refer to van~Dongen's fundamental work\cite{Dongen92}. We calculated with our procedure the ground-state energy on the Bethe lattice with infinite connectivity (bandwidth $W=2\sqrt{2}t^{\ast}$) up to $\mathcal O(t^{12}/U^{-11})$. Taking $t^{\ast}\equiv 1$ as our energy unit in the following, we find that all contributions in the series in $1/U$ vanish, except the first:
\begin{equation}\label{eqn:EnergyFK}
E_0^{\text{FK}}(U)=-\frac{1}{4U}+\mathcal{O}\left(\frac{1}{U^{13}}\right)\,\text{.}
\end{equation}
Next, we verify\cite{ApelPC} our result~(\ref{eqn:EnergyFK}) using the exact solution in ref.~\onlinecite{Dongen92}. We start from the expression of the kinetic energy \mbox{eq.~4.7}; we denote that by $E_0^{\text{FK, T}}(U)$. All ground-state energies are given as densities (intensive thermodynamic variable). We use their spectral representation in order to express the Green function in \mbox{eq.~4.7} in terms of the density of states $z(\epsilon, U)$, \mbox{eq.~7.5} in ref.~\onlinecite{Dongen92}. Thus
\begin{equation}
E_0^{\text{FK, T}}(U)=-2\int\limits_{0}^{\infty}\,\total\epsilon\int\limits_{0}^{\infty}\,\total\epsilon'\,z(\epsilon, U)\,z(\epsilon', U)\,\frac{1}{\epsilon+\epsilon'}\,\text{.}
\end{equation}
Finally, we show by numerical integration for different choices of~$U$ between~$2$ and~$10$ that
\begin{equation}
E_0^{\text{FK, T}}(U)=-\frac{1}{2U}\left(1+\mathcal{O}(10^{-16})\right)
\end{equation}
and that confirms our result, \mbox{eq.~\ref{eqn:EnergyFK}}. (Here, $10^{-16}$ is the numerical accuracy.)

We have to conclude that all higher order hopping contributions to the ground state energy cancel. The reason may be that only one spin species can hop in the Falicov--Kimball model.

\subsection{Hubbard Model}

\begin{figure}
  \includegraphics{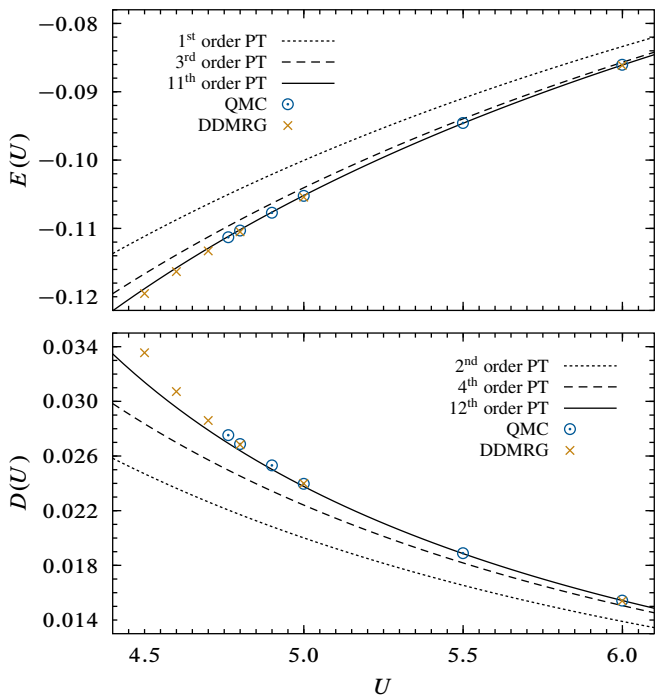}
  \caption{\label{fig:PT_QMC_DDMRG} QMC and DDMRG\cite{Nishimoto04} results for the ground-state energy $E$ (top) and the double occupancy $D$ (bottom) in comparison with PT, see eq.~(\ref{eqn:HubbardEnergy}) and (\ref{eqn:HubbardDouble}). }
\end{figure}

\begin{figure}
  \includegraphics{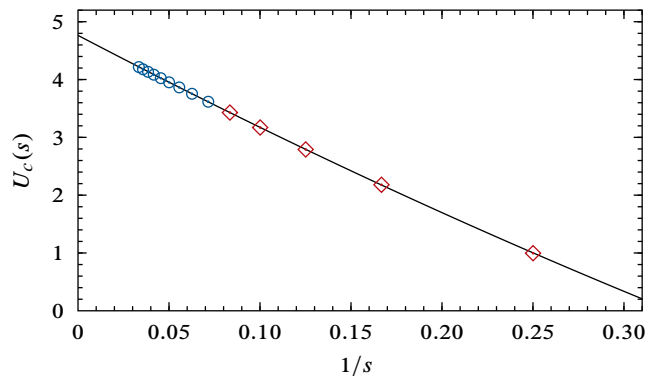}
  \caption{\label{fig:ePT}Construction of ePT: PT values for $U_c(s)=\sqrt{a_{2s+2}/a_{2s}}$ (squares) are extrapolated to $1/s\to 0$ using a quadratic least squares fit (solid line). Evaluations at smaller $1/s$ (circles) define ePT coefficients to all orders, cf.~ref.~\onlinecite{Bluemer04}.}
\end{figure}

\noindent The calculation of the ground-state energy of the Hubbard model to the $11^\text{th}$ order yields
\begin{multline}\label{eqn:HubbardEnergy}
E_0^{\text{H}}(U)=-\frac{1}{2}\frac{1}{U}-\frac{1}{2}\frac{1}{U^3}-\frac{19}{8}\frac{1}{U^5}-\frac{593}{32}\frac{1}{U^7}-\\-\frac{23877}{128}\frac{1}{U^9}-\frac{4496245}{2048}\frac{1}{U^{11}}+\mathcal{O}\left(\frac{1}{U^{13}}\right)\,\text{.}
\end{multline}
Consequently, the double occupancy of the original particles $\frac{1}{L}D(U)=\total E(U)/\total U$ is given by
\begin{multline}\label{eqn:HubbardDouble}
\frac{1}{L}D^{\text{H}}(U)=\frac{1}{2}\frac{1}{U^2}+\frac{3}{2}\frac{1}{U^4}+\frac{95}{8}\frac{1}{U^6}+\frac{4151}{32}\frac{1}{U^8}+\\+\frac{214893}{128}\frac{1}{U^{10}}+\frac{49458695}{2048}\frac{1}{U^{12}}+\mathcal{O}\left(\frac{1}{U^{14}}\right)\,\text{.}
\end{multline}
These perturbation-theoretical (PT) results are shown as solid lines in figure \ref{fig:PT_QMC_DDMRG}. The comparisons of the first (second) and third (fourth) order PT (dotted/dashed lines) demonstrate a fast convergence for the values of $U$ shown. The agreement with QMC (circles) and DDMRG (crosses) results extrapolated to zero temperature is excellent for $U>5$ (smaller than the line width in fig.~\ref{fig:PT_QMC_DDMRG}). As $U$ decreases, devations from these (numerical) DMFT data increase noticeably, since results of finite order PT rapidly become inaccurate as $U$ approaches $U_{c_1}$, the critical interaction. In the following, we describe a method of how to estimate the critical coupling~$U_{c_{1}}$. We assume that the radius of convergence of the $1/U$ expansion of the energy coincides with the critical coupling $U_{c_{1}}$, beyond which the insulating phase becomes stable. We perform an extrapolation of the computer-aided high-order evaluation to infinite order (ePT\cite{Bluemer04}) that exceeds former accuracy in ref.~\onlinecite{Bluemer04}. With $E^\text{H}(U)=\sum_{s=1}^{\infty}a_{2s}U^{1-2s}$, we have $U_{c_{1}}=\lim_{s\to\infty}\sqrt{a_{2s+2}/a_{2s}}$. In figure \ref{fig:ePT}, we plot $\sqrt{a_{2s+2}/a_{2s}}$ against $1/s$. As seen in the figure, the data points are fitted by a nearly straight line as a function of $1/s$. Taking a slight curvature into account in a least-squares fit,
\begin{equation}\label{eqn:ePT}
U_{c_{1}}(s)=\sqrt{\frac{a_{2s+2}}{a_{2s}}}\approx U_{c_{1}}+\frac{u_1}{2s}+\frac{u_2}{(2s)^2}\,\text{,}
\end{equation}
one finds $U_{c_{1}}=4.76$, $u_1= -16.471257$, and $u_2=5.7147072$. The critical exponent (for details see ref.~\onlinecite{Bluemer04}) defined by $E_{\text{crit}}(U)\propto (U-U_{c_{1}})^{\tau-1}$ is obtained with $\tau=3.46$, that gives support to our assumption\cite{Bluemer04} for $\tau=7/2$. 

The ePT estimates for the energy are strongly supported by QMC results\cite{Bluemer04}: $E_{\text{PT}}$ is converged within $\mathcal{O}(10^{-5})$ for $U\geq 6$, while the ePT provides an estimate for $E$ with a precision of the same order above the stability edge of the insulator. These ePT results for $E$ have been reproduced at $U=4.8$, $5.5$, and $6$ within $\mathcal{O}(10^{-6})$ using the self-energy functional approach/dynamical impurity approach (SFT/DIA)\cite{Pozgajcic04}.

Other methods based on DDMRG give $U_{c_{1}}\simeq 4.45$ and $\tau=2.5$ see ref.~\onlinecite{Eastwood03,Nishimoto04}. As seen from figure \ref{fig:PT_QMC_DDMRG}, a high accuracy of data is indispensable for a correct analysis of the transition.

\section*{\MakeUppercase{Conclusions}}

\noindent Using Kohn's unitary transformation, an $1/U$-expansion for the Hubbard model was derived up to order $(1/U)^{11}$ at zero temperature. The expansion has been formulated in terms of diagrammatic rules for the calculation of the ground-state energy and the resulting double occupancy. These rules reduce the calculation of finite-order contributions to an algebra which becomes increasingly complex for higher orders. Any step of the rules is carried out exactly by our computer program. Explicit results were obtained for the ground-state energy up to 11\textsuperscript{th} order in $1/U$.

An inspection of the contributions of the diagrams shows that there are groups of dominant ones, namely the widespread diagrams (first ones in each order in fig.~\ref{fig:ButcherTrees}), and they are significant in view of the metal-insulator transition. This should be analysed quantitatively in the large order limit. Then, even an exact determination of, e.\,g., $U_{c_1}$ might be possible. 

\begin{acknowledgments}
\noindent We thank the referee for many helpful suggestions that greatly improved the presentation of this work and E.~Jeckelmann and---in particular---W.~and V.~Apel for many discussions and help in revising this manuscript.
\end{acknowledgments}

\appendix*
\section{\MakeUppercase{Flowchart of the algorithm}}

\begin{figure*}[p]
  \includegraphics{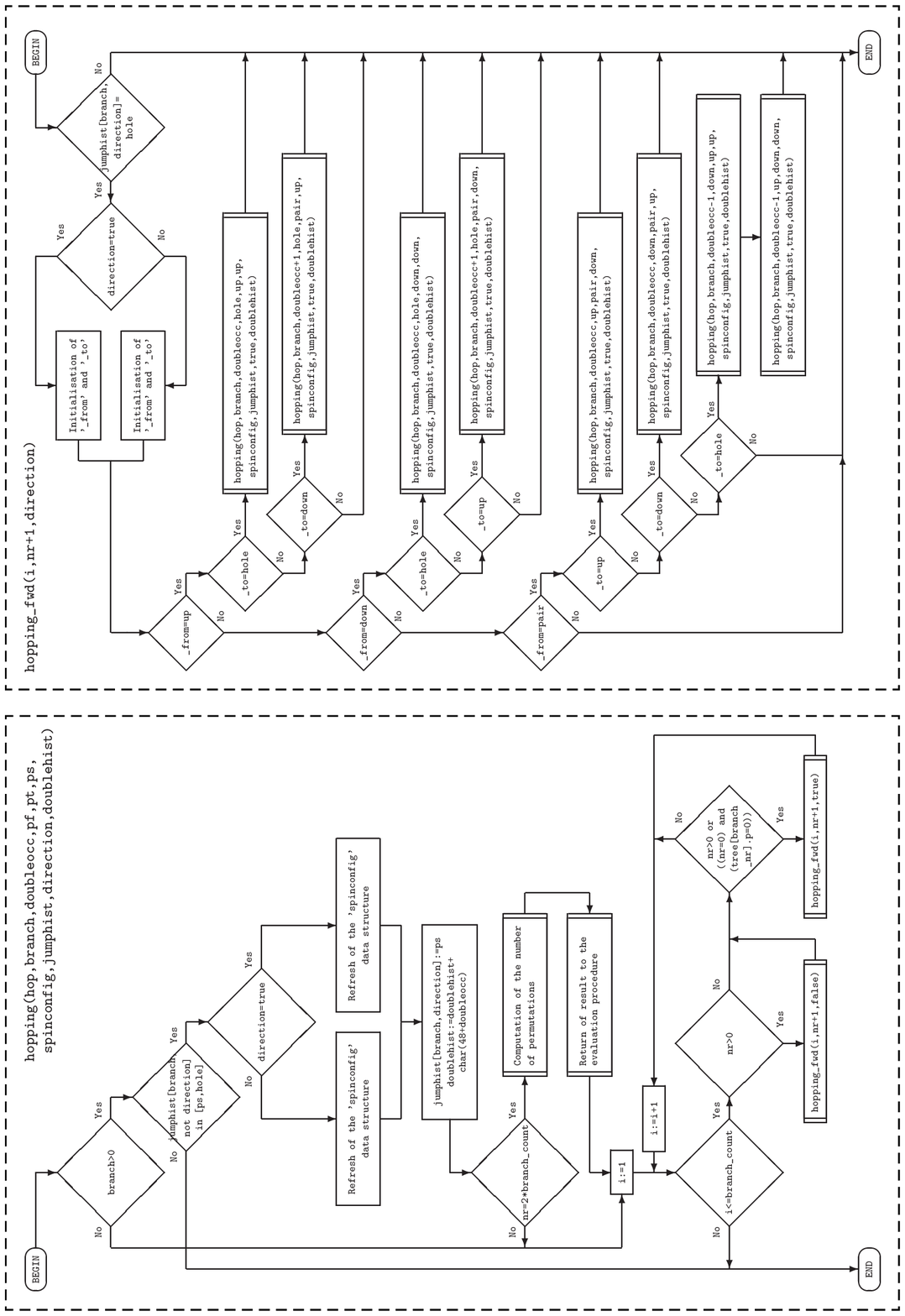}\bigskip\par%
  \caption{\label{fig:flowcharts}Flowcharts of the procedures \texttt{hopping} and \texttt{hopping\_fwd}.}
\end{figure*}

\noindent The kernel of the program is the recursive procedure \texttt{hopping} which is calling the procedure \texttt{hopping\_fwd} and vice versa; their flowcharts are shown in figure \ref{fig:flowcharts}.\bigskip%

{\centering\includegraphics{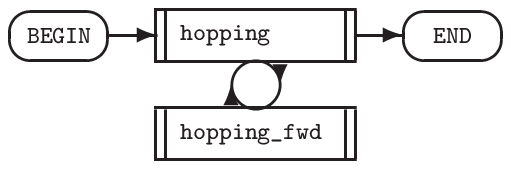}\bigskip\par}

The main task of these ``(non-linear) indirect-recursive'' procedures is to find all electron hopping processes generated by the Hamiltonian, which occur on a set of graphs, as shown in the preceding section. The graphs are defined in such a way that the set of all possible processes of the given order is identical to the set of processes starting from the root of the graphs. All possible spin configurations on the graphs are generated and stored in arrays. Due to symmetry, it is sufficient to occupy the root always by an up-spin leading to $2^{n-1}$ configurations, where $n$ is the number of vertices of the graphs. For the description of the procedures the following variables are used, cf.~fig.~\ref{fig:flowcharts}:
\begin{description}
\item[\texttt{hop}:] number of the current hop
\item[\texttt{branch}:] number of the branch, on which the hop occurs
\item[\texttt{doubleocc}:] number of double occupancies
\item[\texttt{pf}, \texttt{pt}, \texttt{ps}:] spin states (\texttt{hole}, \texttt{up}, \texttt{down}, \texttt{pair}): \texttt{pf} spin state of the site out of which the jump occurs, \texttt{pt} spin state of the site to which the jump occurs, \texttt{ps} jumping spin
\item[\texttt{spinconfig}:] structure describing the current spin configuration on the graph
\item[\texttt{jumphist}:] table containing the jump history on the branches
\item[\texttt{direction}:] direction of the jump, defined by description of the graph: \texttt{true} jump forward, \texttt{false} jump backward
\item[\texttt{doublehist}:] sequence of digits representing the history of the number of double occupancies
\item[\texttt{branch\_count}:] total number of branches in the graph
\end{description}

The initial call of the procedure \texttt{hopping} is done with the following parameters: \texttt{hop=0}, \texttt{branch=0}, \texttt{doubleocc=0}, \texttt{pf=up}, \texttt{pt=down}, \texttt{ps=up}, \texttt{spinconfig}, \texttt{jumphist}, \texttt{true}, \texttt{'0'}. The procedure executes jumps on all branches in both directions; the possibility of a jump is tested through the procedure \texttt{hopping\_fwd}. Its call is done with \texttt{hopping\_fwd(branch\_nr,hop+1,true)} (forwards) or \texttt{hopping\_fwd(branch\_nr,hop+1,false)} (backwards) and it checks if on the current branch a hop has already occured in the given direction. If not, depending on the spin state of the involved sites, the procedure \texttt{hopping} is called, the array of spin states is refreshed, and a next branch is tested. The second jump on a branch has always to be performed by the same spin species as the first one.

The recursion has the property that, in case of exiting the procedure when the jump was not possible, the values of variables resulting from preceding steps are automatically restored.

This construction of the algorithm guarantees that all possible variants of electron jumps are tested in accordance with the assumptions, and that the final spin configuration equals the initial spin configuration; therefore this condition does not need not to be tested. After the last step (carrying the number \texttt{2*branch\_count}), the characteristic factor  
\begin{equation}\label{eqn:fprozess}
f_{\text{proc.}}=(-1)^P\frac{N_{\text{spin}}[\text{proc.}]}{2^{n-1}}\sum_{m}f_{m}\prod_{j=1}^{i}\left(\frac{1}{d_{j}}\right)^{k_j}
\end{equation}
for a process appearing in order $1/U^i$ is computed. In (\ref{eqn:fprozess}) $P$ denotes the number of permuted spin pairs, $N_{\text{spin}}$ is the number of spin configurations not changed by the process. The sum runs over all terms of the hamiltonian which generate the process and $f_{m}$ is the related factor obtained from equation~(\ref{eqn:EffOperator}) and calculated by a separate algorithm. $d_{j}$ are the numbers of double occupancies and $k_{j}$ are also obtained from equation~(\ref{eqn:EffOperator}). The sum of the factors yields the final result.

\bibliographystyle{apsrev}
\bibliography{Algorithm}

\end{document}